
\parskip=\medskipamount
\overfullrule=0pt
\raggedbottom
\def\normalparindent{24pt}
\nopagenumbers
\footline={\ifnum\pageno=1{\hfil}\else{\hfil\rm\folio\hfil}\fi}
\def\endpage{\vfill\eject}
\def\beginlinemode{\endmode\begingroup\parskip=0pt
                   \obeylines\def\\{\par}\def\endmode{\par\endgroup}}
\def\beginparmode{\endmode\begingroup \def\endmode{\par\endgroup}}
\let\endmode=\par
\def\raggedcenter{
                  \leftskip=2em plus 6em \rightskip=\leftskip
                  \parindent=0pt \parfillskip=0pt \spaceskip=.3333em
                  \xspaceskip=.5em\pretolerance=9999 \tolerance=9999
                  \hyphenpenalty=9999 \exhyphenpenalty=9999 }
\def\\{\cr}
\let\rawfootnote=\footnote\def\footnote#1#2{{\parindent=0pt\parskip=0pt
        \rawfootnote{#1}{#2\hfill\vrule height 0pt depth 6pt width 0pt}}}
\def\title{\null\vskip 3pt plus 0.2fill\beginlinemode\raggedcenter\bf}
\def\author{\vskip 3pt plus 0.2fill \beginlinemode\raggedcenter}
\def\affil{\vskip 3pt plus 0.1fill\beginlinemode\raggedcenter\it}
\def\abstract{\vskip 3pt plus 0.3fill \beginparmode{\noindent  ABSTRACT:~}  }
\def\endtitlepage{\endpage\body}
\def\body{\beginparmode\parindent=\normalparindent}
\def\head#1{\par\goodbreak{\immediate\write16{#1}
           {\noindent\bf #1}\par}\nobreak\nobreak}

\def\refto#1{$^{[#1]}$}
\def\ref#1{Ref.~#1}
\def\Ref#1{Ref.~#1}\def\cite#1{{#1}}\def\[#1]{[\cite{#1}]}

\def\(#1){(\call{#1})}
\def\call#1{{#1}}\def\taghead#1{{#1}}
\def\references{\head{REFERENCES}\beginparmode\frenchspacing\parskip=0pt}
\gdef\refis#1{\item{#1.\ }}
\def\endreferences{\body}
\def\endit{\endmode\vfill\supereject}\let\endpaper=\endit
\def\gsim{\mathrel{\raise.3ex\hbox{$>$\kern-.75em\lower1ex\hbox{$\sim$}}}}
\def\lsim{\mathrel{\raise.3ex\hbox{$<$\kern-.75em\lower1ex\hbox{$\sim$}}}}
\def\sla{\raise.15ex\hbox{$/$}\kern-.72em}
\def\iafedir{Instituto de Astronom\'\i a y F\'\i sica del Espacio\\Casilla de
          Correo 67 - Sucursal 28, 1428 Buenos Aires -- Argentina}
\def\fceyn{Departamento de F{\'\i}sica\\
      Facultad de Ciencias Exactas y Naturales, Universidad de Buenos Aires\\
      Ciudad Universitaria - Pabell\'on I, 1428 Buenos Aires, Argentina}
\catcode`@=11
\newcount\r@fcount \r@fcount=0\newcount\r@fcurr
\immediate\newwrite\reffile\newif\ifr@ffile\r@ffilefalse
\def\w@rnwrite#1{\ifr@ffile\immediate\write\reffile{#1}\fi\message{#1}}
\def\writer@f#1>>{}
\def\referencefile{\r@ffiletrue\immediate\openout\reffile=\jobname.ref%
  \def\writer@f##1>>{\ifr@ffile\immediate\write\reffile%
    {\noexpand\refis{##1} = \csname r@fnum##1\endcsname = %
     \expandafter\expandafter\expandafter\strip@t\expandafter%
     \meaning\csname r@ftext\csname r@fnum##1\endcsname\endcsname}\fi}%
  \def\strip@t##1>>{}}

\def\citeall#1{\xdef#1##1{#1{\noexpand\cite{##1}}}}
\def\cite#1{\each@rg\citer@nge{#1}}
\def\each@rg#1#2{{\let\thecsname=#1\expandafter\first@rg#2,\end,}}
\def\first@rg#1,{\thecsname{#1}\apply@rg}
\def\apply@rg#1,{\ifx\end#1\let\next=\relax%
\else,\thecsname{#1}\let\next=\apply@rg\fi\next}%
\def\citer@nge#1{\citedor@nge#1-\end-}
\def\citer@ngeat#1\end-{#1}
\def\citedor@nge#1-#2-{\ifx\end#2\r@featspace#1
  \else\citel@@p{#1}{#2}\citer@ngeat\fi}
\def\citel@@p#1#2{\ifnum#1>#2{\errmessage{Reference range #1-#2\space is bad.}
    \errhelp{If you cite a series of references by the notation M-N, then M and
    N must be integers, and N must be greater than or equal to M.}}\else%
{\count0=#1\count1=#2\advance\count1 by1\relax\expandafter\r@fcite\the\count0,%
  \loop\advance\count0 by1\relax
    \ifnum\count0<\count1,\expandafter\r@fcite\the\count0,%
  \repeat}\fi}
\def\r@featspace#1#2 {\r@fcite#1#2,}    \def\r@fcite#1,{\ifuncit@d{#1}
    \expandafter\gdef\csname r@ftext\number\r@fcount\endcsname%
    {\message{Reference #1 to be supplied.}\writer@f#1>>#1 to be supplied.\par
     }\fi\csname r@fnum#1\endcsname}
\def\ifuncit@d#1{\expandafter\ifx\csname r@fnum#1\endcsname\relax%
\global\advance\r@fcount by1%
\expandafter\xdef\csname r@fnum#1\endcsname{\number\r@fcount}}
\let\r@fis=\refis   \def\refis#1#2#3\par{\ifuncit@d{#1}%
    \w@rnwrite{Reference #1=\number\r@fcount\space is not cited up to now.}\fi%
  \expandafter\gdef\csname r@ftext\csname r@fnum#1\endcsname\endcsname%
  {\writer@f#1>>#2#3\par}}
\def\r@ferr{\endreferences\errmessage{I was expecting to see
\noexpand\endreferences before now;  I have inserted it here.}}
\let\r@ferences=\references
\def\references{\r@ferences\def\endmode{\r@ferr\par\endgroup}}
\let\endr@ferences=\endreferences
\def\endreferences{\r@fcurr=0{\loop\ifnum\r@fcurr<\r@fcount
    \advance\r@fcurr by 1\relax\expandafter\r@fis\expandafter{\number\r@fcurr}%
    \csname r@ftext\number\r@fcurr\endcsname%
  \repeat}\gdef\r@ferr{}\endr@ferences}
\let\r@fend=\endpaper\gdef\endpaper{\ifr@ffile
\immediate\write16{Cross References written on []\jobname.REF.}\fi\r@fend}
\catcode`@=12
\citeall\refto\citeall\ref\citeall\Ref
\catcode`@=11
\newcount\tagnumber\tagnumber=0
\immediate\newwrite\eqnfile\newif\if@qnfile\@qnfilefalse
\def\write@qn#1{}\def\writenew@qn#1{}
\def\w@rnwrite#1{\write@qn{#1}\message{#1}}
\def\@rrwrite#1{\write@qn{#1}\errmessage{#1}}
\def\taghead#1{\gdef\t@ghead{#1}\global\tagnumber=0}
\def\t@ghead{}\expandafter\def\csname @qnnum-3\endcsname
  {{\t@ghead\advance\tagnumber by -3\relax\number\tagnumber}}
\expandafter\def\csname @qnnum-2\endcsname
  {{\t@ghead\advance\tagnumber by -2\relax\number\tagnumber}}
\expandafter\def\csname @qnnum-1\endcsname
  {{\t@ghead\advance\tagnumber by -1\relax\number\tagnumber}}
\expandafter\def\csname @qnnum0\endcsname
  {\t@ghead\number\tagnumber}
\expandafter\def\csname @qnnum+1\endcsname
  {{\t@ghead\advance\tagnumber by 1\relax\number\tagnumber}}
\expandafter\def\csname @qnnum+2\endcsname
  {{\t@ghead\advance\tagnumber by 2\relax\number\tagnumber}}
\expandafter\def\csname @qnnum+3\endcsname
  {{\t@ghead\advance\tagnumber by 3\relax\number\tagnumber}}
\def\equationfile{\@qnfiletrue\immediate\openout\eqnfile=\jobname.eqn%
  \def\write@qn##1{\if@qnfile\immediate\write\eqnfile{##1}\fi}
  \def\writenew@qn##1{\if@qnfile\immediate\write\eqnfile
    {\noexpand\tag{##1} = (\t@ghead\number\tagnumber)}\fi}}
\def\callall#1{\xdef#1##1{#1{\noexpand\call{##1}}}}
\def\call#1{\each@rg\callr@nge{#1}}
\def\each@rg#1#2{{\let\thecsname=#1\expandafter\first@rg#2,\end,}}
\def\first@rg#1,{\thecsname{#1}\apply@rg}
\def\apply@rg#1,{\ifx\end#1\let\next=\relax%
\else,\thecsname{#1}\let\next=\apply@rg\fi\next}
\def\callr@nge#1{\calldor@nge#1-\end-}\def\callr@ngeat#1\end-{#1}
\def\calldor@nge#1-#2-{\ifx\end#2\@qneatspace#1 %
  \else\calll@@p{#1}{#2}\callr@ngeat\fi}
\def\calll@@p#1#2{\ifnum#1>#2{\@rrwrite{Equation range #1-#2\space is bad.}
\errhelp{If you call a series of equations by the notation M-N, then M and
N must be integers, and N must be greater than or equal to M.}}\else%
{\count0=#1\count1=#2\advance\count1 by1\relax\expandafter\@qncall\the\count0,%
  \loop\advance\count0 by1\relax%
    \ifnum\count0<\count1,\expandafter\@qncall\the\count0,  \repeat}\fi}
\def\@qneatspace#1#2 {\@qncall#1#2,}
\def\@qncall#1,{\ifunc@lled{#1}{\def\next{#1}\ifx\next\empty\else
  \w@rnwrite{Equation number \noexpand\(>>#1<<) has not been defined yet.}
  >>#1<<\fi}\else\csname @qnnum#1\endcsname\fi}
\let\eqnono=\eqno\def\eqno(#1){\tag#1}\def\tag#1$${\eqnono(\displayt@g#1 )$$}
\def\aligntag#1\endaligntag  $${\gdef\tag##1\\{&(##1 )\cr}\eqalignno{#1\\}$$
  \gdef\tag##1$${\eqnono(\displayt@g##1 )$$}}
\def\eqalignno#1{\displ@y \tabskip\centering
  \halign to\displaywidth{\hfil$\displaystyle{##}$\tabskip\z@skip
    &$\displaystyle{{}##}$\hfil\tabskip\centering
    &\llap{$\displayt@gpar##$}\tabskip\z@skip\crcr
    #1\crcr}}
\def\displayt@gpar(#1){(\displayt@g#1 )}
\def\displayt@g#1 {\rm\ifunc@lled{#1}\global\advance\tagnumber by1
        {\def\next{#1}\ifx\next\empty\else\expandafter
        \xdef\csname @qnnum#1\endcsname{\t@ghead\number\tagnumber}\fi}%
  \writenew@qn{#1}\t@ghead\number\tagnumber\else
        {\edef\next{\t@ghead\number\tagnumber}%
        \expandafter\ifx\csname @qnnum#1\endcsname\next\else
        \w@rnwrite{Equation \noexpand\tag{#1} is a duplicate number.}\fi}%
  \csname @qnnum#1\endcsname\fi}
\def\ifunc@lled#1{\expandafter\ifx\csname @qnnum#1\endcsname\relax}
\let\@qnend=\end\gdef\end{\if@qnfile
\immediate\write16{Equation numbers written on []\jobname.EQN.}\fi\@qnend}
\catcode`@=12
\magnification=1100
\title{Extrinsic time in quantum cosmology}
\author{S.C. Beluardi$^1$ and Rafael Ferraro$^{1,2}$}
\affil{$^1$\fceyn}
\affil{$^2$\iafedir}
\abstract
An extrinsic time is identified in most  isotropic and homogeneous
cosmological models by matching them with the {\it ideal clock} - a
parametrized system whose only ``degree of freedom'' is  time -.
Once this matching is established, the cosmological models are  quantized in
the
same way as the ideal clock.  The space of  solutions  of the Wheeler-DeWitt
equation is turned out into a Hilbert space by inserting a time dependent
operator in the inner product,  yielding a
unitary theory equivalent to the phase space reduced theory.
\bigskip
PACS:98.80.H

\endtitlepage

\centerline{\bf I - INTRODUCTION}
\vskip 0.5cm
\taghead{1.}
Many efforts have been made in the last two decades to construct a quantum
theory of gravitation . The problems arising when one intends to join, in the
same theory, the principles of Quantum Mechanics and General Relativity, show
that Quantum Gravity is a very complicated and not completely understood
discipline.

One of the most difficult features is the problem of time\refto{aa,ab,ac}. In
Quantum Mechanics, time is an absolute parameter, differently treated from the
other coordinates which turn out to be operators and observables. Instead, in
General Relativity  ``time'' is merely an arbitrary label of a spatial
hypersurface, and physically significant quantities are independent of those
labels: they are invariant under diffeomorfisms.

General Relativity is an example of parametrized system, i.e. a system whose
action is invariant under change of the integrating  parameter
(``reparametrization''). One can obtain such a kind of system by starting from
an action which does not possess the invariance (this means that the
integrating  parameter $t$ {\it is} the time), and
rising the time to the rank of a dynamical variable. So the original
degrees of freedom and the  time are left as functions of some
physically irrelevant parameter $\tau$. The time $t$ can be varied
independently of the other degrees of freedom when a constraint with a Lagrange
multiplier is added.

One of the proposals to understand Quantum Gravity --the so called ``internal
Schr\"odinger interpretation"\refto{aa}-- states that time is hidden among the
dynamical variables, and must  be  picked  up  before quantization.  Once the
time is identified, the system  is  quantized  by  means  of  a Schr\"odinger
equation.  Thus the space of  wave functions can be turned out into a Hilbert
space  by  defining  the usual inner product,  so  providing  a  satisfactory
statistical interpretation.

However, not all the actions which are invariant under reparametrization have
the time hidden among the dynamical variables: Jacobi's
principle\refto{ad,ae} allows to get the trajectory of a conservative system in
the phase space without information about the time evolution, by varying a
parametrized action which does not contain the time among the dynamical
variables.\refto{hj}

Therefore a criterion is being needed to establish whether the  time is
hidden in a parametrized system, joined with a method to pick it up.
The action for a parametrized system has the general form:
$$S[q^i,p_i,N]=\int_{\tau_1}^{\tau_2}\left(p_i{dq^i\over d\tau} - N{\cal H}
(q^i,p_i)\right)\ d\tau, \eqno(11)$$
where ${\cal H}$ is the constraint and $N$ is the Lagrange multiplier.

This action is invariant under reparametrization:
$$\delta q^i=\epsilon (\tau) {dq^i\over d\tau}(\tau),\quad\quad \delta
p_i(\tau)
=\epsilon (\tau){dp_i\over d\tau}(\tau),\quad\quad \delta N={d\over d\tau}(N
\epsilon),\quad\quad \epsilon (\tau_1)=0=\epsilon (\tau_2),\eqno(12)$$
which is equivalent to change $\tau\rightarrow\tau + \epsilon (\tau)$ on the
path $(q(\tau),p(\tau))$, $\int_{\tau_1}^{\tau_2} Nd\tau$ remaining invariant.
Besides the action \(11) is invariant under a gauge transformation:
$$\delta q^i=\epsilon (\tau) \{q^i,{\cal H}\},\quad\quad \delta p_i(\tau)
=\epsilon (\tau)\{p_i,{\cal H}\},\quad\quad \delta N={d\epsilon\over
d\tau},\quad\quad \epsilon (\tau_1)=0=\epsilon (\tau_2).\eqno(13)$$
Both transformations are not independent but differ by an ``equation-of-motion
symmetry''\refto{aj}. On the classical path, the reparametrization is equal to
a gauge transformation with parameter $N\epsilon$.

The solutions of the Hamilton equations associated with \(11) have the form:
$$ q^i=q^i\left(\int N d\tau\right),\quad\quad p_i=p_i \left(\int N d\tau
\right)  .\eqno(14)$$
So $\int Nd\tau$, instead of $\tau$, plays the role of physically meaningful
time parameter.

It is always possible to locally solve \(14) for $\int Nd\tau$,
i.e. there are locally well defined functions $t=t(q^i,p_i)$ in the phase space
coinciding with $\int Nd\tau$ when evaluated on the functions
\(14). However a global solution may not exist; if it does exist, one calls it
a ``global phase time"\refto{af,ag,ah,ai}.  To find a global phase time
amounts  to get a globally well defined function $t=t(q^i,p_i)$ such that
$$\{\ t,\ {\cal H}\}\vert_{{\cal H}=0}\ =\ 1.\eqno(15)$$
In  fact,  $dt/d\tau = N(\tau) \{\ t,\ {\cal H}\}\vert_{{\cal H}=0} = N(\tau)$
on any trajectory.  Furthermore, let us  suppose that we know a globally well
defined function $\bar t=\bar t(q^i,p_i)$ such that
$$\{\ \bar t,\ {\cal H}\}\vert_{{\cal H}=0}\ =\ F(q,p) > 0.\eqno(16)$$
Then $\bar t$ is a global phase time associated with the constraint
$$\bar{\cal H} \equiv F(q,p) {\cal H}.\eqno(17)$$
But $\bar{\cal H}$ and $\cal H$ are
entirely equivalent;    they  describe  the same parametrized system, because
their respective Hamiltonian  vectors
--$\ {\bf H}\ =\ \left({\rm H}^q,{\rm H}^p\right)$ $=\
\left({\partial {\cal H}\over \partial p},
-{\partial {\cal  H}\over  \partial q}\right)\ $--, whose field lines coincide
with the  classical trajectories, are proportional on the constraint surface
(remember that the $\tau$ evolution is physically irrelevant).
Therefore $\bar t$ should  be  also  consider  as a global phase time for the
system described by $\cal H$. This teachs us that global phase time
should be rather defined  by  means  of \refto{af}
$$\{\ t,\ {\cal H}\}\vert_{{\cal H}=0}\ >\ 0.\eqno(18)$$
Eq.\(18) tells us that
$$ {\rm H}^A\ {\partial t\over \partial x^A} \ > \ 0 ,$$
where $x^A = (q, p)$,  i.e.    $t(q,p)$  monotonically increases along any
dynamic trajectory;  each surface $t=constant$  is crossed by the dynamic
trajectories only once (so the ${\bf H}$ field lines are necessarily open).
\bigskip

A  global phase time can play the role  of  time  in  an
internal Scr\"odinger interpretation.  The quantization should be performed by
solving the constraint  for  $p_t$  --the momentum associated with time--,
to obtain  the  Hamiltonian   entering  the
Schr\"odinger equation for the ``reduced" system.\refto{aa,ad}

Unfortunately the solution of eq.\(18),  whenever a global one exists, is not
unique.  This lack of uniqueness is  called  the ``multiple  choice  problem"
\refto{aa}, because it can lead to different quantum theories.

Even if  the  job  of  solving $p_t$ was successful for some chosen time, the
Hamiltonian of the  reduced  system may result so intricate as to desist from
quantizing the system.

\bigskip

Another  proposal to understand  Quantum  Gravity  consists  in  solving  the
Wheeler-DeWitt equation, which  comes  from  constraining the wave
function according to the Dirac method
$$\hat{\cal H}\ \varphi\ =\ 0,\eqno(19)$$
where $\hat{\cal H}$ is an operator  associated  with the constraint.
This is a second order hyperbolic differential equation in the
usual variables of  gravity, which resembles the Klein-Gordon equation.  Like
this,  the Wheeler-DeWitt equation allows for a conserved inner  product,
defined on a ``spacelike" hypersurface in the ``superspace"; however
this product fails to be positive definite, so leaving  this approach without
a clear interpretation.\refto{al}

\bigskip

In this paper we shall search for global phase time in cosmological
models. We are going to deal with an ``extrinsic time" \refto{aa,ak}, i.e.
one which is intended to be associated not only with the coordinates but
also with  the  momenta.  Extrinsic times deserve special attention in Quantum
Gravity, because it
is known that there  is  no  chance  of  reducing the  system by
identifying  a  time  among  the  coordinates  --like   in  the  case  of  the
relativistic particle--,  due to the fact that the potential
in the superhamiltonian constraint is not definite positive \refto{aa,al}.

A  simple model of parametrized system
--the ``ideal clock"-- will suggest us how  the space of solutions of the
Wheeler-DeWitt equation can be turned out into  a  Hilbert  space,  with
an  inner product matching the expectation values of the
Schr\"odinger approach associated with the extrinsic time.

In section II the ideal clock --a parametrized system having no genuine degrees
of freedom-- is introduced.   It  is  quantized  by  following the Dirac
recipe,
where a singular operator must be inserted in the inner product, in order to
recover the physical expectation values of the corresponding reduced system.
In section III the ideal clock is suitably
generalized, in order to allow a comparison with the cosmological
models.  In section IV  isotropic and homogeneous
cosmological models are studied. An extrinsic  time is
picked  up  in  most  of  these minisuperspaces.  Finally, in section  V  the
problem  of quantizing a parametrized system with genuine degrees of freedom,
namely a homogeneous scalar field in a Robertson-Walker metric, is glanced
in light of the experience acquired with the ideal clock.
\vskip 1cm
\centerline{\bf II - THE IDEAL CLOCK}
\vskip 0.5cm
\taghead{2.}
Let us start by considering a system without degrees of freedom. Then, its
``action'' is not a functional of dynamical variables, but merely an arbitrary
function of the  time $t$:
$$ S = \int f(t)\ dt .$$
In order to parametrize the system one changes the integration variable $t$ to
$\tau$:
$$S=\int f(t){dt\over d\tau}\ d\tau .$$
Thus $S$ could be regarded as a Hamiltonian functional action by
identifying $f(t)$ with $p_t$. However, in order that the function $S$ does not
change, this identification should reenter through a constraint:
$${\cal H} = p_{t} - f(t) = 0, \eqno(21)$$
and the action is:
$$ S[t,p_{t},N]  =  \int  d\tau  \left[  p_{t}  {dt\over  d\tau}  -  N  {\cal
H}\right].\eqno(22)$$
Now $(t,p_t)$ is a pair of dynamical conjugated variables. We remark that the
constraint is linear in $p_t$; this will play a significant role in the
quantization of the system.

By varying \(22) with respect to $t, p_t$ and $N$, one obtains the dynamic
equations and the constraint:
$$\eqalign{{dt\over d\tau}& = N,\cr {dp_{t} \over d\tau}& = N{df\over dt},\cr
{\cal H}& = p_{t} - f(t) = 0.\cr}$$
{}From the first equation:
$${dt\over d\tau}= N\quad\Rightarrow\quad t = \int N\ d\tau,$$
so $N$ relates the variations of $\tau$ with the variations of $t$, and it is
known as {\it lapse function}. Then the second equation reads:
$${d\over d\tau}\left(p_t - f(t)\right)=0,$$
so $p_t - f(t)$ is a constant of motion. Finally, the third equation fixes the
conserved quantity.
As the only generalized coordinate of this system is the  time $t$, we
call it {\it ideal clock}.

We can proceed with the quantization of this system. The wave function
satisfies the Schr\"odinger equation
$$i{d\psi\over d\tau}=N \hat {\cal H} \psi .$$
But, following the Dirac method, the constraint must be imposed on the wave
function,
$$\hat {\cal H} \psi = \left[\hat {p_t} - f(t)\right]\psi = 0,$$
to get the ``physical states''. So the physical states do not depend on $\tau$.
Since $\hat {p_t}=-i\partial /\partial t$, the constraint equation turns out
to be
$$i{\partial \psi\over \partial t} = -\ f(t)\ \psi ,\eqno(23)$$
which is  the  Schr\"odinger  equation  in the time $t$ for the {\it reduced}
system described by the Hamiltonian $h = -f(t)$; its only solution is
$$\psi (t)\ = \ e^{i\int f(t)dt}.\eqno(24)$$
This wave function is nothing but a phase, which can be understood remembering
that the clock has not degrees of freedom.

Let us define the inner product as:
$$<\psi , \psi>=\int_a^b dt\ \psi^{\ast}(t)\psi (t) ,$$
$\hat p_t=-i\partial /\partial t$ is an hermitian operator on the space of
physical states \(24) whatever $(a,b)$ is (in fact, $\psi^{\ast}\psi
\mid_a^b=0)$. Since time is an unbounded variable, the interval should be
$(-\infty ,\infty) $ or $(a,\infty) $. So the inner product $<\psi ,\psi >$
between
physical states diverges, as it is typical for constrained systems. In order to
get a physically meaningful result $(\psi ,\psi )$, a hermitian singular
operator must be inserted\refto{aj}:
$$(\psi ,\psi )=<\psi ,\hat{\mu}_{t_o}\psi >, \eqno(25)$$
$$\hat {\mu}_{t_o}=\delta(\chi_{t_o}(t))\mid \{\chi_{t_o},{\cal H}\}\mid ,$$
where $\chi_{t_o}$ is any function such that $\chi_{t_o}(t)=0$ is a gauge
condition $(\{\chi,{\cal H}\}\not= 0)$
fixing $t$ in $t_o$ (for instance $\chi_{t_o}=t - t_o$), and $\mid\{\chi,{\cal
H}\}\mid$ is the Fadeev-Popov
``determinant''. In this way, the physical state \(24) becomes normalized:
$$(\psi,\psi)=1.$$
In eq.\(25) $t_o$ must be regarded as the time at which the physical inner
product
is evaluated. Since $(\psi,\psi)$ does not depend on $t_o$, the time evolution
is unitary. In particular
$$\hat t(\hat \mu_{t_o} \psi)=t_o\hat\mu_{t_o}\psi.$$
\bigskip
We remark the importance of having in ${\cal H}$ a term linear in the momentum,
in order to reach the Schr\"odinger equation \(23) in the time $t$.
However,
that is not the case in  parametrized systems of physical interest like
General Relativity, where the constraint is quadratic in the momenta.
In order to change the constraint of the ideal clock to one being quadratic in
the momentum, let us choose the (arbitrary) function $f(t)$ to be:
$$f(t) = t^2 ,$$
and perform the canonical transformation:
$$Q=p_t ,\qquad\qquad P=-t .$$
Thus the constraint of the ideal clock results in
$${\cal H}= -P^2\ +\ Q ,\eqno(27)$$
and the ``dynamics'' of the ideal clock can be got by varying the action
$$S = \int\left[P{dQ\over d\tau} - N\left(-P^2 + Q\right)\right]d\tau ,$$
which differs from the original action in a surface term.

The global phase time is:
$$t(Q,P)=-P$$
which, of course, fulfills eq.\(15). The constraint surface ${\cal H}=0$ is a
parabola in variables $(Q,P)$, which intersects once and only once each surface
$t=constant$. The
Hamiltonian vector
$${\bf H}=\left(-2P\ ,\ -1\right)$$
crosses the $t=constant$ surfaces in the direction of increasing time.

The physical quantum states for the system with constraint \(27) come from the
solution of the equation
$${d^2\varphi\over dQ^2} + Q\varphi=0 .\eqno(28)$$
Differently from \(23), this is a second order equation which has two
linearly independent solutions: the Airy functions $Ai(-Q),\ Bi(-Q)$. However,
there are boundary conditions to be fulfilled. If $Q \in$
$(-\infty,\infty)$, $Bi(-Q)$ should be discarded because
diverges when $Q\rightarrow -\infty$ \refto{q}. Then
$$\varphi (Q)=\sqrt{2\pi}\ Ai(-Q)\eqno(29)$$
which goes to zero in both infinities. The physical state \(29) is the Fourier
transform of the wave function \(24). One can guess what the singular operator
$\hat\mu_{t_o}$ should be in the $Q$-representation by inserting the identity
twice in \(25):
$$\hat\mu_{t_o}\varphi (Q)={1\over 2\pi}\int_{-\infty}^{\infty} dQ'\ e^{-it_o(Q
- Q')}\ \varphi (Q')\eqno(210) .$$
Then
$$(\varphi,\varphi)=<\varphi,\hat\mu_{t_o}\varphi>=1 ,$$
since
$$\left|\int_{-\infty}^{\infty} dQ\ e^{-it_oQ}\ Ai(Q)\right| =1 .$$
Besides
$$-\hat P(\hat\mu_{t_o}\varphi)=t_o(\hat\mu_{t_o}\varphi) ,$$
in agreement with the fact that $-P$ is the time.
\vskip 1cm
\centerline{\bf III - GENERALIZATION OF THE QUADRATIC CONSTRAINT}
\vskip 0.5cm
\taghead{3.}
The constraint \(27) is still too simple  to fit those of the
cosmological models. One can get a more general form for the constraint of the
ideal clock, although keeping it quadratic in the momentum, by performing
another canonical transformation --a coordinate change-- to the
variables $(\Omega,\pi_{\Omega})$ defined as:
$$Q\ =\ V(\Omega),\quad\quad\quad P\ =\ {d\Omega\over dQ}\ \pi_{\Omega}\ =\
\left(dV\over d\Omega\right)^{-1}\ \pi_\Omega,\eqno(31)$$
where $V(\Omega)$ is a monotonous function unbounded from above\refto{q}.
In these variables, the constraint turns out to be:
$${\cal H}(\Omega,\pi_{\Omega})\ =\ -\left(dV\over d\Omega\right)^{-2}\
\pi_{\Omega}^2\ +\ V(\Omega) ,\eqno(32)$$
and the global phase time is:
$$t(\Omega,\pi_{\Omega})\ =\ -P(\Omega,\pi_{\Omega})\ =\ -\left(dV
\over d\Omega\right)^{-1}\ \pi_\Omega .\eqno(33)$$
Since the coordinate change \(31) should not modify the wave function (whenever
the wave function is regarded as a scalar), then the kinetic term of the
constraint operator should be associated with the invariant D'alembertian
operator.
\vskip 0.5cm
In general, the constraints appearing in problems of interest have the form:
$$\bar {\cal H}(\Omega,\pi_{\Omega})  = g(\Omega)\ \pi_{\Omega}^2 + v(\Omega)
,\eqno(34)$$
with $g(\Omega)< 0$. So we are faced to the issue of knowing whether a
constraint like \(34) hides or not an ideal clock.

The constraint \(34) describes an ideal clock if it has the form \(17), where
$\cal H$ should be the one of eq.\(32). Then
$$V\ V'^2=-{v\over g} .\eqno(35)$$
Since $V(\Omega)$ in eq.\(31) is a monotonous function, $V(\Omega)$ has no
more than one zero. Because of eq.\(35), the same must be accomplished by the
function $v(\Omega)$ (in other
case, $\bar {\cal H}$ does not correspond with an ideal clock). Let $\Omega_0$
be such that $v(\Omega_0)=0$; then the solution of eq.\(35) is
$$V(\Omega) = sign(v)\ \left({3\over 2}\int_{\Omega_0}^{\Omega} \sqrt{{\mid
v\mid\over -g}}\ d\Omega\right)^{2/3} ,\eqno(36)$$
and
$$F\ =\ {v\over V}\  =\ - g\ V'^2 .\eqno(37)$$
It might happen that $F(\Omega)$ is zero or ill-defined in $\Omega_0$.  Since
${\bf\bar H}\vert_{{\cal H}=0}$ $=F {\bf H}$, in such a case the system would
stop or its evolution would result ill-defined, when $Q_0\equiv V(\Omega_0)=0$
is  reached, ie at $t(Q_0)$ $= -P(Q_0)$  $=0$.  So, in order that $t$ $= - P$
can  evolve  till  infinity,  the  time  should  be  regarded  as a  positive
variable, and only values  $t_o >0$  should be considered in  the  gauge
condition.
\bigskip
Since the constraints $\cal H$ and $\bar {\cal H}=F {\cal H}$ are equivalent,
they should lead to the same quantization. In fact the relation between both
Dirac quantizations is as follows:
$$\eqalignno{\hat{\bar{\cal H}}&  = \hat F^{1/2}\ \hat {\cal H}\ \hat F^{1/2} ,
\cr \bar\varphi &=\hat F^{-1/2}\ \varphi , &(311)\cr
\hat {\bar\mu}&= \hat F^{1/2}\ \hat\mu\ \hat F^{1/2} . \cr}$$
Then
$$\eqalign{\hat{\bar {\cal H}}\bar \varphi &=0 ,\cr (\bar\varphi,\bar\varphi )
&= (\varphi,\varphi) .\cr}$$
The transformation \(311) can be regarded as induced by a finite unitary
transformation  $U\ =\ \exp \left[(i/2) (\eta \ln F\ {\cal P} - {\cal P}
\ln F\  \eta)\right]$,  in  the  Hilbert  space  of  the  $BRST$ quantization
where  $(\eta,{\cal  P})$    are   the  ghosts  associated  with  the
constraint (cf \Ref{aj,fs}).
\medskip
In the next sections we shall examine in what extent the Hamiltonian constraint
of the isotropic and homogeneous cosmological models correspond with an ideal
clock.

\vskip 1cm
\centerline{\bf IV - COSMOLOGICAL MODELS}
\vskip 0.5cm
\taghead{4.}
The action of General Relativity with cosmological constant $\tilde\Lambda$ is:
$$S=-{1\over 16\pi G}\int dt\ d^3x \sqrt{-g}\left({\cal R} + 2\tilde\Lambda
\right) ,$$
where ${\cal R}$ is the curvature scalar. We shall restrict ourselves to
spatially
isotropic and homogeneous geometries, so we shall deal with
Robertson-Walker metrics:
$$ds^2={\tilde N}^2(\tau)\ d\tau^2\ -\ R(\tau)^2\left({dr^2\over 1-Kr^2}\ +\
r^2 d\theta^2\ +\ r^2 sin^2\theta d\varphi^2\right) ,$$
with $K=\pm 1,0$. In this way the original system is left with a finite number
of degrees of freedom ({\it minisuperspace models}).

The Robertson-Walker metrics fulfill the
requirement for the consistency of the minisuperspace models, viz the same
dynamics must result either by replacing the minisuperspace metric in the
Einstein equations or by varying the action with respect to the remaining
degrees of freedom:
$$S=-{3\pi\over 4G}\int {\tilde N}d\tau\left({R\dot R^2\over {\tilde N}^2}- KR
+ {\tilde\Lambda\over 3}R^3\right) .$$
The Hamiltonian form of the action is:
$$S = \int d\tau \left(\dot\Omega \pi_{\Omega} - N \bar {\cal H}\right) ,$$
where:
$$\bar {\cal H} = -{1\over 4}e^{-3\Omega}\ \pi_{\Omega}^2
- Ke^{\Omega} + \Lambda e^{3\Omega} ,\eqno(41)$$
and:
$$\Omega\ =\ \ln\left[{\left({3\pi\over 4G}\right)^{1/2} R}\right],\quad\quad
N\ =\ \left({3\pi\over 4G}\right)^{1/2} \tilde N,\quad\quad
\Lambda\ =\ \left({3\pi\over 4G}\right)^{-1}\ {\tilde\Lambda\over 3}.
\eqno(42)$$
So
$$g(\Omega) = -{1\over 4}e^{-3\Omega},\qquad\qquad v(\Omega) = -Ke^{\Omega}
+ \Lambda e^{3\Omega}.\eqno(43)$$

We are going to look for a global phase time in the minisuperspaces resulting
of the combinations of $\Lambda$ and $K$ which make sense in the constraint
equation. Thus the cases $(\Lambda < 0\ ;\ K = 0,1)$ and $(\Lambda = 0\ ;
\ K = 1)$ are excluded.

In the Minkowski case $(K=0=\Lambda)$, the constraint \(41) is satisfied only
if $\pi_{\Omega}=0$; the Hamiltonian vector is null in $\pi_{\Omega}=0$: the
system does not evolve. Then this case is not an ideal clock.

Almost all of the remaining cases will prove to be ideal clocks.
According to the method described in the former section, the potential
$V(\Omega)$ is:
$$V(\Omega) = sign(v)\ \left[3\int_{\Omega_0}^{\Omega}e^{2\Omega}\sqrt{\left|
-K\ +\ \Lambda\ e^{2\Omega}\right|}\ d\Omega \right]^{2/3} .\eqno(44)$$

For the cases $(\Lambda\geq 0;K=-1)$ and $(\Lambda>0;K=0)$ it is
$v(\Omega)\ >\ 0\qquad \forall\ \Omega$;
so the lower boundary in \(44) should be $-\infty$ (thus
$V\rightarrow 0$ when $\Omega\rightarrow -\infty$ like $-v/g$, in according
with eq.\(35)). When $sign(\Lambda)\ =\ sign(K)$ it is
$$\Omega_0={1\over 2}\ln\left({K\over \Lambda}\right) .$$
\vskip 0.5cm
What follows is a summary of the results for each minisuperspace:
\vskip 0.5cm
1) $\Lambda=0$ ; $K=-1$

$$\eqalign{Q&= V(\Omega) = \left({3\over 2}\right)^{2/3}e^{4\Omega /3},\qquad
Q\ \epsilon\ (0,\infty),\cr F&= \left({3\over 2}\right)^{-2/3} e^{-\Omega /3}=
\sqrt{{2\over 3}}\ Q^{-1/4} .\cr}$$
The global phase time is:
$$t(\Omega,\pi_{\Omega}) = -P = -{1\over 2}\left({3\over 2}\right)^{-1/3}
e^{-(4/3)\Omega}\ \ \pi_{\Omega} .$$
$t_o$ should be taken to be positive $(P<0)$, because $F$ is not well
behaved at $Q=0$.
\vskip 0.5cm
2) $\Lambda > 0\ ;\ K = 0,-1$

$$Q=V(\Omega) = \Lambda^{-2/3}\left[K + \left(\Lambda e^{2\Omega} - K\right)^{
3/2}\right]^{2/3},\qquad Q\ \epsilon\ (0,\infty) ,$$
$$\eqalign{ F &= \Lambda^{2/3} e^{\Omega}\left[1 + K\left(\Lambda e^{2\Omega}
-K\right)^{-3/2}\right]^{-2/3}=\cr &=\Lambda^{-1/2}Q^{-1}\left(\Lambda Q^{3/2}
- K\right)^{2/3}\left[K + \left(\Lambda Q^{3/2} - K\right)^{2/3}\right]^{1/2}
,\cr}$$
and
$$t(\Omega,\pi_{\Omega})=-P=-{1\over 2}\Lambda^{-1/3}\left[1 + K\left(\Lambda
e^{2\Omega} - K\right)^{-3/2}\right]^{1/3} e^{-2\Omega}\ \pi_{\Omega} .$$
$t_o$ should be positive, because $F$ is not well behaved at $Q=0$.
\vskip 0.5cm
3) ($\Lambda>0$ ; $K=1$) , ($\Lambda<0$ ; $K = -1$)
$$\eqalign{Q&=V(\Omega)=|\Lambda|^{-2/3}\left(\Lambda e^{2\Omega}  - K\right)
,\cr F &=|\Lambda|^{2/3} e^{\Omega} = \left(\Lambda Q +
|\Lambda|^{1/3}\right)^{1/2} .\cr}$$
In the case $(\Lambda<0;K=-1)$ the potential $V(\Omega)$ is bounded from above;
therefore this case is not an ideal clock. In the case $(\Lambda>0;K=1)$ is $Q\
\epsilon\ (-\Lambda^{-2/3},\infty)$ and the global phase time is:
$$t(\Omega,\pi_{\Omega}) = -P = -{1\over 2}\Lambda^{-1/3} e^{-2\Omega}\
\pi_{\Omega},\qquad -\infty < t_o < \infty .$$
\bigskip
As we can see, it is possible  to  match several cosmological models with the
ideal clock by means of an extrinsic time.   We remark that an extrinsic time
was already introduced by York \refto{ak} in the
context of superspace  for  closed  3-manifolds without cosmological constant.
The York  time  is $t_{York} =$ $2/3 \gamma^{-1/2} \gamma^{ab} \pi_{ab}$;  it
is canonically conjugated to minus the volume scale: $p_{t_{York}}$
$=-\gamma^{1/2}$.  On
an hypersurface $t_{York} =$ constant, the  Hamiltonian constraint determines
the  scale  factor  $\gamma$,  once  the traceless part $p_{ab}$ of momenta
$\pi_{ab}$ satisfying the supermomentum constraints, and the unimodular metric
$\sigma_{ab} =$ $\gamma^{-1/3}\gamma_{ab}$ --conjugated to $p_{ab}$--  are
given  on  the hypersurface.
So  $t_{York}$  and $\sigma_{ab}$  are  good  dynamical  variables,  and  the
Hamiltonian of the system is
$h = -p_{t_{York}} = \gamma^{1/2}(\sigma_{ab}, p_{ab}, t_{York})$\refto{ak,aa}.

The homogeneous and isotropic minisuperspaces  studied  in  this section, have
the scale factor as their  only  dynamical variable;
$\sigma_{ab}$ and $p_{ab}$ are frozen, and
no such a geometry exists in the superspace considered by
York ($\Lambda = 0$, $K =1$ does not make sense in the constraint coming from
\(41)).   So no comparison is possible with York's time.

\vskip 1cm
\centerline{\bf V - GENUINE DEGREES OF FREEDOM}
\vskip 0.5cm
\taghead{5.}
So far we have studied too simple models: they have no degrees of freedom,
they are nothing but time. The following step should be the inclusion of
genuine  degrees of freedom, like a matter field. In order to glance how
an extrinsic time should be managed in such a case,
let us consider a homogeneous scalar field minimally coupled  to  a
Robertson-Walker geometry  with  $K  =  0$,  $\Lambda > 0$.  The action for a
scalar field $\tilde\phi$ is
$$S_{matter}\ =\ {1\over 2}\int dt\ d^3x \sqrt{-g}\left(g^{\mu\nu}
\tilde\phi_{;\mu}\tilde\phi_{;\nu} - \tilde m^2\tilde\phi^2\right),$$
which turns out to be
$$S_{matter}\ =\ \int {\tilde N}\ dt\ \pi^2\ R^3\ \left({\dot
{\tilde\phi^2}\over {\tilde N}^2} - {\tilde m}^2\tilde\phi^2\right),$$
in the minisuperspace under consideration.

The Hamiltonian form of this action is
$$S_{matter}\ =\ \int d\tau \left[\dot\phi\ \pi_{\phi} + \dot\Omega\
\pi_{\Omega} - N\left({1\over 4}e^{-3\Omega}\ \pi_{\phi}^2 +
m^2 e^{3\Omega}\phi^2\right)\right],$$
where
$$\phi\ =\ \pi\left({3\pi \over 4G}\right)^{-1/2} \tilde\phi, \quad\quad
m\ =\ \left({3\pi\over 4G}\right)^{-1/2}\tilde m.$$
The constraint for the entire system $S_{grav} + S_{matter}$ is
$$\bar {\cal H} = {1\over 4}e^{-3\Omega}\left(\pi_{\phi}^2 - \pi_{\Omega}^2
\right) + m^2 e^{3\Omega}\phi^2 + \Lambda e^{3\Omega}\eqno(51)$$
For simplicity, let us choose $m = 0$. Using the variables
$$t\ =\ -P\ =\ -{1\over 2} \Lambda^{-1/3}\ e^{-2\Omega}\ \pi_{\Omega},\ \ \
\ \ \ \ \ \ \ \ \ p_t\ =\ Q\ =\ \Lambda^{1/3}\ e^{2 \Omega}\ >\ 0,\eqno(52)$$
the constraint becomes
$$\bar{\cal H}\  =\ \Lambda^{1/2} p_t^{-3/2}\left(p_t^3 - t^2 p_t^2 + {1\over
4} \pi_\phi^2\right).\eqno(53)$$
We want to check that $t$ is still a time:
$$\{t,\bar{\cal  H}\}\vert_{\bar{\cal  H}=0}\  =\    \Lambda^{1/2}  p_t^{-1/2}
\left(3 p_t - 2 t^2\right).$$
So, in order that $t$  remains  a  time, it should be granted that $3 p_t$ $> 2
t^2$ on all the dynamic trajectories. By introducing the variables
$$y \ \equiv\ {3\ p_t\over 2\  t^2}\ >\ 0,\quad\quad\quad
b\ \equiv\ {27\ \pi_\phi^2\over 16\ t^2}\ \geq\ 0,$$
the constraint is written as
$$\bar{\cal H}\ =\ {4\over 27} \Lambda^{1/2} t^6  p_t^{-3/2} \left(2 y^3 - 3
y^2 + b\right).$$
Thus the dynamics  lies  on  the  positive  roots  of  a  cubic  polynomial
parametrized by $b$.
This polynomial has a maximum at $y_{max} = 0$, and  a  minimum at $y_{min} =
1$. One of its roots is $y_1(b) \leq 0$. If $b \leq 1$ there are
two positive roots:  $y_2(b) < 1 < y_3(b)$  ($b < 1$), and $y_2(b) = 1 =
y_3(b)$  ($b = 1$).  When $\pi_\phi = 0$ (ie,
$b=0$) it is $y_1=0=y_2$,  $y_3=3/2$.  Therefore the dynamics lies in the root
$y_3$, because $y_3 = 3/2$ means $p_t = t^2$, which is the constraint equation
of the former sections. So let us concentrate on the trajectories satisfying $y
= y_3(b)$.  As was said,  $t$  is  a  time only if $y > 1$;  since $y = y_3
\geq 1$, we  shall  only  prove  that  $y$  never  reachs  $1$  along  a
dynamic trajectory, by showing that $y$ increases when $b$ is near to $1$:
$$\{y,\bar{\cal H}\}\vert_{\bar{\cal H} = 0}\ =\ {4\over 27} \Lambda^{1/2} t^6
p_t^{-3/2}\ \{y,b\}\vert_{\bar{\cal  H}=0}\ =\ {4\over 3} \Lambda^{1/2} t^3 b\
p_t^{-3/2}$$
Like in the  case  without  matter, $t$ should be intended to be positive.  In
fact, when $t$ goes  to zero, then also $p_t$  goes to zero
(because $1\leq y=y_3\leq 3/2$); so we would be faced with a singular point of
the Hamiltonian vector, where the dynamics is ill-defined.
Then $y$ increases when $b\not= 0$,  and stops when $b=0$ at the
value  $y=y_3(b=0)=3/2$  \refto{mas}. Thus $y$ remains bigger than $1$, so
proving that $t$ is a  time  if  the
constraint  is  fulfilled  through  the  root  $y_3(b)$.  The constraint
$y = y_3(b)$ means
$${\cal H}\ \equiv \ p_t - {2\over 3} t^2 y_3(b)\ =\ 0.\eqno(54)$$
The function $h(t, \pi_\phi) \equiv - {2\over 3} t^2 y_3(b)$ is the
Hamiltonian for the reduced  system (associated with the chosen time $t$),  so
one could quantize the matter field by means of a Schr\"odinger equation,
$$\hat{\cal H} \ \psi(\phi,t)\ =\ 0,$$
achieving a unitary theory with the usual inner product:
$$(\psi_1,\psi_2)\ =\ \int d\phi\ \psi_1^*(\phi,t_o)\ \psi_2(\phi,t_o)\ =\
\int dt\ d\phi\ \psi_1^*\ \delta (t-t_o)\ \psi_2,\quad t_o>0.$$
Unfortunately the root $y_3$ is a so complicated function of $b$ as to desist
from solving the Schr\"odinger equation.
\bigskip
As  was  shown, the internal Schr\"odinger interpretation
can be connected with the solutions of the Wheeler-DeWitt  equation
$$\hat{\bar{\cal
H}}(\Omega,-i{\partial\over\partial\Omega},\phi,-i{\partial\over\partial\phi})\
\bar\varphi\ = \ 0,$$
where $\hat{\bar{\cal H}}$  is  an  operator  associated  with  the constraint
\(51).
This equation is  much more tractable than the Schr\"odinger equation for the
reduced system.    The  problem  here  is  how the space of solutions obeying
suitable boundary conditions, is turned out into a Hilbert space. In sections
II and III we solved this issue at light of the trivial ideal clock:
the physical inner product requires the insertion of a singular  operator
$\hat{\bar\mu}_{t_o}$:
$$(\bar\varphi_1,\bar\varphi_2)\ =
<\bar\varphi_1,\hat{\bar\mu}_{t_o}\  \bar\varphi_2>$$
In order  to know the singular operator, we should factorize out the physical
root in the constraint \(53), in such a way that
$$\bar{\cal H} = F(t,p_t,\phi,\pi_\phi) {\cal H},$$
where $\cal H$ is the one of eq.\(54). Then
$$\hat{\bar\mu}_{t_o}\ =\ \hat F^{1/2}\ \hat\mu_{t_o}\ \hat F^{1/2},$$
where $\hat\mu_{t_o}$ is the  one  of  eq.\(210), which results from applying
the                canonical        transformation        $(t,p_t)\rightarrow
\left(Q(\Omega),P(\Omega,\pi_\Omega)\right)$  to    the    insertion  $\delta
(t-t_o)$.

\medskip
This completes the  scheme  of  quantization  based  on  the solutions of the
Wheeler-DeWitt equation and an extrinsic time. In spite of appearances,
difficulties are still present. In fact, the operator
$\hat  F$  is  complicated  enough  as  to  prevent  from  computing  the
probabilities.

\vskip 1cm
\centerline{\bf ACKNOWLEDGEMENT}
\vskip 0.5cm
We wish to thank E.Calzetta and M.Henneaux for helpful conversations.
\vskip 1cm

\references

\vskip 0.5cm
\refis{aa} K.V.Kucha\v r, in {\it Proceedings of the 4th. Canadian Conference
on General Relativity \& Relativistic Astrophysics}, eds. G.Kunstatter,
D.Vincent and J.Williams, World Scientific, (1992).

\refis{ab} K.V.Kucha\v r, in {\it Conceptual Problems of Quantum Gravity}, eds.
A.Ashtekar and J.Stachel,\goodbreak  Birkh\"auser, Boston, (1991).

\refis{ac} M.Henneaux and C.Teitelboim, Phys.Lett. B {\bf 222}, 195 (1989);
W.Unruh and  R.Wald,  Phys.Rev.    D  {\bf  40}, 2598 (1989);  K.V.Kucha\v r,
Phys.Rev. D {\bf 43}, 3332 (1991).

\refis{ad} C.Lanczos, {\it The Variational Principle of Mechanics}, Dover
(1986).

\refis{ae} J.D.Brown and J.W.York, Phys.Rev. D {\bf 40}, 3312 (1989).

\refis{af} P.H\'aj\'\i cek, Phys.Rev. D {\bf 34}, 1040 (1986).

\refis{ag} P.H\'aj\'\i cek and M.Sch\"on, Class. Quantum Grav. {\bf 7}, 861
(1990).

\refis{ah} P.H\'aj\'\i cek, Class. Quantum Grav. {\bf 7}, 871 (1990).

\refis{ai} P.H\'aj\'\i cek, J.Math.Phys. {\bf 30}, 2488 (1989).

\refis{aj} M.Henneaux and C.Teitelboim, {\it Quantization of Gauge Systems},
Princeton University Press, New Jersey, (1992).

\refis{ak} J.W.York, Phys.Rev.Lett. {\bf 28}, 1082 (1972).

\refis{al}  K.V.Kucha\v  r,  in  {\it  Quantum  Gravity 2:  A  Second  Oxford
Symposium}, eds.   C.J.Isham,  R.Penrose  and  D.W.Sciama,  Clarendon  Press,
Oxford, (1981).

\refis{hj} The point at Jacobi's variational principle is  the building of an
action where the information about the time interval $t_f  - t_i$ is replaced
by  the  knowledge  of  the  energy, which enters the Jacobi's  action  as  a
property of the system.

\refis{q} $Q$ should be always intended to be unbounded from above, since this
requirement warrants that $t=-P$ reachs the infinity on the classical
trajectory $P^2=Q$. But there is not obstacle to regard $Q$ as belonging to an
interval $(A,\infty)$, rather than $(-\infty,\infty)$. In such a case one
should select the wave function fulfilling $\varphi (A)=0$, in order that
$\hat P=-i\partial /\partial Q$ be hermitian.

\refis{fs} R.Ferraro and D.Sforza, in preparation.

\refis{mas} In the case $m\not= 0$, one can still prove that $y$ never reachs
$1$ if $\Lambda^{-1} m^2 < 12$.

\endreferences

\end